\documentclass[12pt]{article}
\usepackage{amsmath,amssymb}
\usepackage{graphicx,color}
\numberwithin{equation}{section}
\usepackage{cite}
\usepackage{bm}
\usepackage{dcolumn}
\newcommand{\be}{\begin{equation}}
\newcommand{\bea}{\begin{eqnarray}}
\newcommand{\eea}{\end{eqnarray}}
\newcommand{\ba}{\begin{array}}
\newcommand{\ea}{\end{array}}
\newcommand{\ee}{\end{equation}}

\expandafter\ifx\csname mathbbm\endcsname\relax

\else

\fi \textheight 22cm \textwidth 15cm \topmargin 1mm
\def\N{{\cal \nu}}
\begin{document}
\begin{titlepage}
\hfill
\vbox{
    \halign{#\hfil         \cr
           IPM/P-2007/077 \cr
                      } % end of \halign
      }  % end of \vbox
\vspace*{20mm}
\begin{center}
{\Large {\bf On 5D Small Black Holes}\\
}

\vspace*{15mm}
\vspace*{1mm}
{Mohsen Alishahiha$^a$, Farhad Ardalan$^{a,b}$, Hajar Ebrahim$^a$ \\  and \\  Subir Mukhopadhyay$^c$}

 \vspace*{1cm}

{\it ${}^a$ Institute for Studies in Theoretical Physics
and Mathematics (IPM)\\
P.O. Box 19395-5531, Tehran, Iran \\ }

\vspace*{.4cm}

{\it ${}^b$ Department of Physics, Sharif University of Technology \\
P.O. Box 11365-9161, Tehran, Iran}

\vspace*{.4cm}

{\it ${}^c$ Institute of Physics, Bhubaneswar 751~005, India.}

\vspace*{2cm}
%%\maketitle
\end{center}

\begin{abstract}
Using higher order corrections we argue that five
dimensional ${\cal N}=2$ and ${\cal N}=4$ small black holes exhibit supersymmetry enhancement in near horizon
geometry leading to eight and sixteen supercharges, respectively. Using this enhancement
at supergravity level we can identify the global supergroup of the near
horizon geometry. In particular we show how this supergroup distinguishes between
small and large black holes in ${\cal N}=2$ case.  

\end{abstract}
\end{titlepage}

\section{Introduction}

In this paper we study symmetry of the near horizon geometry of the extremal black holes in ${\cal N}=2$ and $4$ supergravities in five dimensions.
The corresponding black holes could be either small or large depending on whether the corresponding classical 
horizon area is zero or non-zero, respectively.
In the ${\cal N}=2$ case both small and large black holes are $\frac{1}{2}$ BPS. But in the ${\cal N}=4$ 
case the large black hole is $\frac{1}{4}$ BPS 
whereas the small one is $\frac{1}{2}$ BPS.

An interesting feature of these extremal black holes is that in the near horizon limit they 
exhibit supersymmetry enhancement \cite{Cham}. More precisely, at the leading order, in the 
near horizon limit large black holes
undergo supersymmetry doubling, i.e. the near horizon of large black holes in both ${\cal N}=4$ and 
${\cal N}=2$ preserves eight supercharges. 
On the other hand the small black holes in both cases are singular 
at the leading order and as a result going near horizon 
we will not led to supersymmetry doubling, i.e. for ${\cal N}=2$ small black holes at near core limit
there are just four supercharges while for ${\cal N}=4$ case the number of supercharges are eight. The same 
goes for large and small black strings.

In this paper we will show that taking into account the $R^2$ corrections 
the small black holes will also exhibit supersymmetry
doubling\footnote{Note however, that although our explicit computations are given for a specific higher
order correction \cite{Hanaki:2006pj}, the argument can be
extended to any higher order corrections as long as the near horizon geometry remains $AdS_2\times S^3$.}. 
That means in ${\cal N}=4$ case the small black holes preserve all the sixteen supercharges while for the 
${\cal N}=2$ theory the near horizon geometry of small black holes preserve eight supercharges. 

The main property underlying the supersymmetry enhancement is the appearance of $AdS_2$ 
geometry in the near horizon limit due to the extremality. This is the case both for large and 
small black holes when the higher order corrections are taken into account. 
Actually the main motivation of the present paper is the fact that the higher order corrections
can stretch the horizon of the extremal small black holes leading to the 
non-singular $AdS_2\times S^{d-2}$ near horizon geometry \cite{sen2}. 

Having established the supersymmetry enhancement for extremal small black holes, it 
may be possible to study the near horizon
symmetry which in turn help understanding of  the holographic dual of string theory/gravity on 
the $AdS_2$ background.
The point is that the ${\rm AdS}_{2}/{\rm CFT}_{1}$ correspondence is not well-understood 
in contrast to the higher dimensional cases (see for example 
\cite{{Strominger:1998yg},{Gaiotto:2004pc},{Gaiotto:2004ij},{Azeyanagi:2007bj}}). 
Our study might shed a new light on this subject. Note also that gravity on $AdS_2$ geometry is important on
its own as it carries entropy unlike the higher dimensional cases. The $AdS_2$ space is a background which naturally appears in the general near horizon geometry of the extremal black holes,
and therefore its holographic dual could be used to understand the entropy of the black holes better.

The extremal black holes in five dimensional ${\cal N}=2$ supergravity in the presence of 
supersymmetrized $R^2$ corrections were studied in \cite{{Castro:2007sd},{Castro:2007hc},{Alishahiha:2007nn},{Castro:2007ci},{Cvitan:2007en}} where 
it was shown that these corrections stretch the horizon leading to $AdS_2\times S^3$ 
near horizon geometry. In the present work we will obtain the symmetry of the near horizon geometry of
these black holes. In particular we will see that this symmetry can distinguish between the small and
large black holes in the ${\cal N}=2$ theory where both have the same number of eight supercharges 
in the near horizon limit.

Our main results are that in ${\cal N}=4$ theory the near horizon geometry of small black holes preserve 
sixteen supercharges and the corresponding global near horizon symmetry is $OSp(4^*|4)\times SU(2)$.
In the ${\cal N}=2$ case the near horizon geometry of the small black hole preserves eight supercharges
with global near horizon symmetry of $OSp(4^*|2)\times SU(2)$. 
This is to be compared with the symmetry of the near horizon geometry of the large black holes
$SU(1,1|2)\times SU(2)$ \cite{Townsend}.

The paper is organized as follows. In section two we will introduce our notation. In section three we show how 
the $R^2$ corrections lead to supersymmetry enhancement for small black holes in ${\cal N}=2$ five dimensional 
supergravity. In section four we will study the supersymmetry enhancement for small black holes in ${\cal N}=4$ model 
where we will also obtain the near horizon symmetry. In section five, using the global near horizon geometry of
${\cal N}=4$, we will show how the near horizon symmetry distinguishes between small and large black holes  in ${\cal N}=2$ model.
The last section is devoted to discussions and conclusions.  

\section{${\cal N}=2$ 5D Black holes}

In this section 
we will begin with the  study of both small and large black holes which are half BPS and therefore 
preserve 4 supercharges
in ${\cal N}=2$ five dimensional supergravity.
We will then take up the supersymmetry behavior of the solutions in the near horizon limit. We will
do this at the leading order as well as up to $R^2$ corrections.
In the leading order this question has been addressed in \cite{{Cham},{Townsend},{Sabra}}. 
  
\subsection{Basic setup}

We will now briefly review the result of \cite{Hanaki:2006pj} to set our notation.
To study ${\cal N}=2$ supergravity in five
dimensions in the presence of $R^2$ corrections, the authors of \cite{Hanaki:2006pj} apply
the superconformal formalism \cite{{Kugo:2000hn},{Fujita:2001kv},{Bergshoeff:2001hc},{Bergshoeff:2004kh}}.
In this approach a five dimensional theory which is invariant under a
larger group, i.e. superconformal group is taken as an initial point  and  by imposing a gauge fixing condition 
the conformal supergravity is reduced to the standard supergravity model. 

The representation of superconformal group includes Weyl, vector and
hyper multiples. The 
bosonic part of the Weyl multiplet contains the vielbein $e_\mu^a$, a two-form auxiliary
field $v_{ab}$, and a scalar auxiliary field $D$. The bosonic part of the vector multiplet 
contains  one-form gauge fields $A^I$ and scalar fields $X^I$, where $I=1,\cdots,n_v$ labels
generators of the gauge group. The hypermultiplet contains scalar fields ${\cal A}_\alpha^i$,
where $i=1,2$  is the $SU(2)$ doublet index and $\alpha=1,\cdots, 2n$ refers to $USp(2n)$ group.
Although we choose not to couple the theory to matter, the hypermultiplet is used to gauge fix the dilatational symmetry 
which reduces the action to the standard ${\cal N}=2$ supergravity action. 

In this notation at leading order the bosonic part of the action is\cite{Hanaki:2006pj}
\be
I=\frac{1}{16\pi G_5}\int d^5x{\cal L}_0,
\ee
with
\bea
{\cal L}_0&=&\partial_a{\cal A}_\alpha^r\partial^a{\cal A}^\alpha_r+
(2\N+{\cal A}^2)\frac{D}{4}+(2\N-3{\cal A}^2)\frac{R}{8}+(6\N-{\cal A}^2)\frac{v^2}{2}
+2\N_IF_{ab}^Iv^{ab}
\cr &+&\frac{1}{4}\N_{IJ}(F^I_{ab}F^{J\;ab}+2\partial_a X^I\partial^a X^J)
+\frac{g^{-1}}{24}C_{IJK}\epsilon^{abcde}A^I_a F^J_{bc}F^K_{de},
\label{5acts}
\eea
where ${\cal A}^2={\cal A}_{\alpha\;ab}^r{\cal A}^{\alpha\;ab}_r,\;v^2=v_{ab}v^{ab}$ and
\be
\N=\frac{1}{6}C_{IJK}X^IX^JX^K,\;\;\;\;\;\;\;\N_I=\frac{1}{2}C_{IJK}X^JX^K,\;\;\;\;\;\;\;
\N_{IJ}=C_{IJK}X^K.
\ee

The supersymmetrized higher order action with four-derivative terms has been obtained 
in \cite{Hanaki:2006pj} using the superconformal formalism. Of course in what follows the explicit form of the
higher order action is not needed. The only thing that we need is the supersymmetry variations of 
the fields.

The supersymmetry variations of the fermions in Weyl, vector and hyper
multiplets (taking only the bosoinc terms) are\footnote{Here the covariant curvature ${\hat R}^{ij}_{\mu\nu}$ is defined by ${\hat R}^{ij}_{\mu\nu}=2\partial_{[\mu}
V^{ij}_{\nu]}-2V^{i}_{[\mu\;k}V^{kj}_{\nu]}+{\rm fermionic\; terms}$, where $V^{ij}_{\mu}$ is a boson
in the Weyl multiplet which is in ${\bf 3}$ of the $SU(2)$. We note, however, that for the
solutions we are going to consider this term vanishes.} 
\bea
\delta \psi_\mu^i &=&{\cal D}_\mu\varepsilon^i+\frac{1}{2}v^{ab}\gamma_{\mu ab}\varepsilon^i-\gamma_\mu\eta^i,\cr
\delta \chi^i&=&D\varepsilon^i-2\gamma^c\gamma^{ab}{\hat {\cal D}}_av_{bc} \varepsilon^i
+\gamma^{ab}{\hat R}_{ab}(V)^i_j\varepsilon^j
-2\gamma^a\varepsilon^i
\epsilon_{abcde}v^{bc}v^{de}+4\gamma\cdot v\eta^i,\cr
\delta\Omega^{I\;i}&=&-\frac{1}{4}\gamma\cdot F^I\varepsilon^i-\frac{1}{2}\gamma^a\partial_aX^I\varepsilon^i-X^I\eta^i,\cr
\delta\zeta^\alpha&=&\gamma^a\partial_a{\cal A}^\alpha_i\varepsilon^i-\gamma\cdot v\varepsilon^i{\cal A}^\alpha_i+3{\cal A}^\alpha_i
\eta^i,
\label{SUSY}
\eea
where garavitino $\psi_\mu^i$ and the auxiliary Majorana spinor $\chi^i$ come from the Weyl multiplet, while
the gaugino $\Omega^{I\;i}$ and $\zeta^\alpha$ come from  vector and hypermultiplets, respectively.
Here $i=1,2$ is $SU(2)$ doublet index.

The covariant derivatives of spinors are defined by\footnote{In this notation 
$\gamma^{a_1a_2\cdots a_n}=\frac{1}{n!}\gamma^{[a_1}\gamma^{a_2}\cdots \gamma^{a_n]}$.}
\be
{\cal D}_\mu=\partial_\mu+\frac{1}{4}\omega_\mu^{ab}\gamma_{ab},
\ee
where $\omega_\mu^{ab}$ are the spin connection one forms related to the vielbein through the Cartan equation
\be
de^a+\omega^a_b\wedge e^b=0.
\ee

To fix the gauge it is convenient to set ${\cal A}^2=-2$. In this gauge using the last equation
of (\ref{SUSY}) one can express $\eta^i$ in terms of $\varepsilon^i$. Plugging this into the other 
supersymmetry variations one gets
\bea
\delta \psi_\mu^i &=&{\cal D}_\mu\varepsilon^i+\frac{1}{2}v^{ab}\gamma_{\mu ab}\varepsilon^i-
\frac{1}{3}\gamma_\mu\gamma^{ab}v_{ab}\varepsilon^i,\cr
\delta \chi^i& =&D\varepsilon^i-2\gamma^c\gamma^{ab}{\hat {\cal D}}_av_{bc} \varepsilon^i-2\gamma^a\varepsilon^i
\epsilon_{abcde}v^{bc}v^{de}+\frac{4}{3}(\gamma^{ab} v_{ab})^2\varepsilon^i,\cr
\delta\Omega^{I\;i}&=&-\frac{1}{4}\gamma\cdot F^I\varepsilon^i-\frac{1}{2}\gamma^a\partial_aX^I\varepsilon^i
-\frac{1}{3}X^I\gamma^{ab}v_{ab}\varepsilon^i,
\label{SUSY2}
\eea

\subsection{Black hole solutions}

The ${\cal N}=2$ five dimensional supergravity model considered here is known to have
several black hole/string solutions. Here we will only consider the black hole solutions.
The black holes could be either large or small depending on whether at leading order they have
non-vanishing or vanishing horizon. In what follows, for simplicity, we will consider a model with
three vector multiplets (STU model), though the results can be generalized to any other models.

\subsubsection{Leading order}

At leading order in the gauge of ${\cal A}^2=-2$ one can integrate out the auxiliary fields $D$ and $v$ by making use 
of their equations of motion. From the equations of motion of the auxiliary fields one finds
\be 
\nu=\frac{1}{6}C_{IJK}X^IX^JX^K=1,\;\;\;\;\;\;\;\;v_{ab}=-\frac{3}{4}X_IF^I_{ab},
\ee 
where $X_I=\frac{1}{6}C_{IJK}X^IX^JX^K$. Therefore the leading order action reads
\be
{\cal L}_0=R -\frac{1}{2}G_{IJ}F^I_{ab}F^{Jab}-
{\cal G}_{ij} \partial_a \phi^i\partial^a \phi^j+\frac{g^{-1}}{24}\epsilon^{abcde}C_{IJK}
F_{ab}^IF_{cd}^JA^K_e.
\label{5act}
\ee
The parameters in the action (\ref{5act}) are defined by
\be
G_{IJ}=-\frac{1}{2}\partial_I\partial_J\log {\cal \nu}|_{{\cal \nu}=1}\;,\;\;\;\;\;\;\;\;\;\;\;\;
{\cal G}_{ij}=G_{IJ}\;\partial_iX^I\partial_jX^J|_{{\cal \nu}=1}\;,
\label{5actd}
\ee
where $\partial_i$ refers to a partial derivative with respect to the scalar fields $\phi^i$.
In fact doing this, we recover the very special geometry underlying the theory in the leading order.

The black hole solution of the above action has the following form\footnote{For simplicity we will
consider $STU=1$ model, though the procedure is the same for other cases.} \cite{{Cham},{Sabra}}
\bea
ds^2&=&e^{-4U}dt^2-e^{2U}(dr^2+r^2d\Omega_3^2),\;\;\;\;\;\;e^{2U}X_I=\frac{1}{3}H_I,\cr
F^I_{tr}&=&-\partial_r (e^{-2U} X^I),\;\;\;\;\;\;\;\;\;\;\;\;\;\;\;\;\;\;\;\;\;\;\;\;\;\;\;\;e^{6U}=H_1H_2H_3,
\label{bh5}
\eea
where $H_I=h_I+\frac{q_I}{r^2}$.

Using the supersymmetry variation one can see that the above solution preserves four supercharges constrained by 
\cite{{Cham},{Sabra}}
\be
\gamma^{{\hat t}}\varepsilon^i=-\varepsilon^i.
\ee 

Now the aim is to study the supersymmetry properties of the near horizon limit of the black hole solution (\ref{bh5}). To do
this we recognize two different cases. The first case is when $q_I\neq 0$ for all $I$ where we get a large black hole with
non-vanishing horizon and therefore non-zero macroscopic entropy is given by $S_{BH}=2\pi\sqrt{q_1q_2q_3}$. One may also consider 
the case where one of the charges is zero, say $q_1=0$. This corresponds to a small black hole with vanishing
horizon and macroscopic entropy to this order.

For the large black hole the near horizon geometry is given by
\be
ds^2=l^2\left[(\rho^2dt^2-\frac{d\rho^2}{\rho^2})-4d\Omega_3^2\right],
\ee
where $l=(q_1q_2q_3)^{1/6}/2$ and $r^2=(q_1q_2q_2)^{1/2} \rho/2$. 

Using the supersymmetry variation expressions at leading order it was shown in \cite{{Cham},{Sabra}}
that the above near horizon solution 
preserves all the eight supercharges. Thus supersymmetry enhancement occurs in the near horizon limit
of five dimensional large black holes.

For the small black hole where $q_1=0$ the near horizon geometry can be recast in the following form
\be
ds^2=l^2\left[(\rho^4dt^2-\frac{d\rho^2}{\rho})-\frac{16}{9}\rho d\Omega_3^2\right].
\ee
This solution is singular and unlike the large black hole preserves only four supercharges, which
can easily be verified using the above near horizon geometry and the supersymmetry variations (\ref{SUSY2}). 
The simplest
way to find the number of supercharges is to note that the near horizon geometry of the small black hole at leading order
is indeed a specific example of the general solution (\ref{bh5}) which preserves four supercharges. Therefore taking
near horizon geometry of the small black hole does not lead to supersymmetry enhancement. As we shall see the higher order
corrections will change the situation.
  
\subsubsection{$R^2$ corrections}

Four derivative corrections to the 5D ${\cal N}=2$ supergravity have been obtained in \cite{Hanaki:2006pj}
and the resultant corrected black hole and black string solutions have been extensively studied
recently in \cite{{Castro:2007sd},{Castro:2007hc},
{Alishahiha:2007nn},{Castro:2007ci}} where the corresponding corrections to their entropy have also been
obtained.

The results of these papers are mainly based on the near horizon information and the assumption that
the near horizon geometry is either $AdS_2\times S^3$ or $AdS_3\times S^2$. Of course ultimately
one would like to construct a supergravity solution in the presence of $R^2$ terms which
interpolates between asymptotic flat and these near horizon geometries. Having found the
corrected near horizon geometry does not necessarily mean that there is a solution with the above
near horizon geometry. However, an explicit solution, based on numerical computations,
has been presented in \cite{Castro:2007hc}.
 
An important feature of adding higher order corrections is that these terms resolve the
singularity of the small black hole and black string solutions, leading to $AdS_2$ and $AdS_3$ near 
horizon geometry, respectively.
In view of the supersymmetry enhancement for the large black hole of the previous section, it would be
natural to ask whether taking into account the $R^2$ corrections would lead to
supersymmetry enhancement for small black holes too.

The corrected near horizon geometry of the small black hole is given by \cite{{Castro:2007hc},
{Alishahiha:2007nn}}
\be
ds^2=l^2\left[(r^2dt^2-\frac{dr^2}{r^2})-4d\Omega_3^2\right],\;\;\;\;\;v_{\hat{t}\hat{r}}=\frac{3}{4}l.
\label{ads2R2}
\ee
where $l=\frac{1}{6} \sqrt{\frac{c}{8}q_2q_3}$. The scalars in the vector multiplet are constant and in 
what follows we do not need their explicit form. $d\Omega_3^2$ is the round metric on a three-sphere of 
unit radius and can be written in Hopf coordinates
\be
d\Omega_3^2 = d\theta^2 + \sin^2{\theta}~ d\phi^2 + \cos^2{\theta}~ d\psi^2 .
\ee 
The components of vielbein are
\begin{gather}
e^{\hat t} = lr dt,
\quad
e^{\hat r} = \frac{l}{r}dr,
\quad
e^{\hat\theta} = 2l d\theta
\quad
e^{\hat\phi} = 2l\sin{\theta} d\phi
\quad
e^{\hat\psi} = 2l\cos{\theta} d\psi.
\label{vierbein}
\end{gather}
The components of the inverse vierbein are given by
\begin{gather}
e^{{\hat t}t}=\frac{1}{lr}
\quad
e^{{\hat r}r}=-\frac{r}{l}
\quad
e^{{\hat\theta}\theta}=-\frac{1}{2l}
\quad
e^{{\hat\phi}\phi}= - \frac{1}{2l\sin{\theta}}
\quad
e^{{\hat\psi}\psi}= - \frac{1}{2l\cos{\theta}}.
\label{inversevierbein}
\end{gather}
Since in the present case all components of vierbein are diagonal the expressions for components of spin connection reduce to
\begin{gather}
(\omega_\mu)^{{\hat\mu}{\hat\nu}} = 
- (\omega_\mu)^{{\hat\nu}{\hat\mu}} = 
e^{{\hat\nu}\nu}\partial_\nu e^{\hat\mu}_{~\mu}
\label{spinconnection1}\end{gather}
Substituting (\ref{vierbein}) and (\ref{inversevierbein}) in 
(\ref{spinconnection1}) we see only non-zero components of spin connection are
\begin{gather}
(\omega_t)^{{\hat t}{\hat r}} = -r ,
\quad
(\omega_\phi)^{{\hat\phi}{\hat\theta}} = - \cos{\theta},
\quad
(\omega_\psi)^{{\hat\psi}{\hat\theta}} =  \sin{\theta},
\end{gather}
and the ones obtained by permuting the hatted indices.
All other components of spin connection vanish. 

Using the near horizon solution (\ref{ads2R2}) the gravitino variation in (\ref{SUSY2}) reads 
\begin{gather}
\delta\psi_\mu^i = (\partial_\mu + \frac{1}{4}(\omega_\mu)^{ab}\gamma_{ab} + v_{{\hat r}{\hat t}} 
\gamma_\mu^{~~{\hat r}{\hat t}} - \frac{2}{3} v_{{\hat r}{\hat t}}\gamma_\mu\gamma^{{\hat r}{\hat t}} ) 
\varepsilon^i,
\end{gather}
and setting the variations of components of gravitino equal to zero leads to the following set of equations:
\bea
\left(\partial_t -\frac{r}{2} \gamma^{\hat{r}}(\gamma^{\hat{t}}+1)\right) \varepsilon^i &=& 0, \label{t}\\
\left(\partial_r + \frac{1}{2r}\gamma^{\hat t} \right)\varepsilon^i &=& 0 ,\label{r} \\
\left(\partial_\theta +\frac{1}{2} \gamma^{{\hat\theta}{\hat r}{\hat t}}\right)\varepsilon^i &=& 0 ,\label{eta} \\
\left(\partial_\phi - \frac{1}{2} \cos{\theta} \gamma^{{\hat\phi}{\hat\theta}}
+ \frac{1}{2} \sin{\theta} \gamma^{{\hat\phi}{\hat r}{\hat t}} \right) \varepsilon^i &=& 0 , \label{phi}\\
\left(\partial_\psi + \frac{1}{2} \sin{\theta} \gamma^{{\hat\psi}{\hat\theta}}
+ \frac{1}{2} \cos{\theta} \gamma^{{\hat\psi}{\hat r}{\hat t}} \right) \varepsilon^i &=& 0 .\label{psi}
\eea
It is then easy to show that the following solves all the above equations\footnote{We note that there are
two other equations coming from the variation of $\Omega^{I\;i}$ and $\chi$. Nevertheless it can be shown that 
these to equations do not impose any further constraint on the Killing spinors.}
\bea
\varepsilon^i=\sqrt{\frac{r}{l}} \Omega\;\varepsilon_0^i,\;\;\;\;\;\;
\lambda^i=\frac{l}{2}(t-\frac{\gamma^{\hat{r}}}{r})\;\varepsilon^i,\;\;\;\;\;\;{\rm with}\;\;\;
\Omega=e^{\frac{1}{2}\gamma^{{\hat t}\hat{r}\hat{\theta}}\theta}\;  
e^{-\frac{1}{2}\gamma^{\hat{\theta}\hat{\phi}}\phi} \; 
e^{\frac{1}{2}\gamma^{{\hat t}\hat{r}\hat{\psi}}\psi}
\eea
where $\varepsilon_0^i$ is a constant spinor such that $\gamma^{\hat{t}}\varepsilon_0^i=-\varepsilon_0^i$. Moreover
two different chiralities are related by $\gamma^{\hat{r}}$, i.e. $\varepsilon^i_-=\gamma^{\hat{r}}\varepsilon^i_+$.
Since the five dimensional theory is non-chiral, one may choose $\gamma^{\hat{t}\hat{r}\hat{\theta}\hat{\phi}\hat{\psi}}=1$,
therefore the angular dependence of the spinors may be simplified as follows
\be
\Omega=\sqrt{\frac{r}{l}}\;e^{\frac{1}{2}\gamma^{{\hat \psi}\hat{\phi}}\theta}\;  
e^{-\frac{1}{2}\gamma^{\hat{\theta}\hat{\phi}}(\psi+\phi)}. 
\ee
In conclusion we note that altogether there are eight supercharges in the near horizon limit for the small
black hole when $R^2$ correction is taken into account, where the solution is non-singular with
$AdS_2\times S^3$ near horizon geometry. 

We note also that these Killing spinors are exactly the same as that found \cite{{Sabra},{Townsend}} where the authors 
found the same Killing spinors for near horizon geometry of large black holes at leading order. Here, however, 
we have seen that when $R^2$ corrections are taken into account, the same result can also be applied to 
small black holes whose near horizon geometry is $AdS_2\times S^3$ in the presence of $R^2$ corrections.

This might be understood as follows. The only parameter which
appears in the expressions of the supersymmetry variation is the value of the central charge 
evaluated at near horizon. This is also the
parameter which fixes the radii of AdS and sphere factors. Therefore one may always rescale the coordinates such 
that central charge can be dropped from the supersymmetry variations. On the other hand, when taking into account
the higher order corrections ($R^2$ correction in our case) with the assumption of $AdS_2\times S^3$ near horizon
geometry, the corrections will only change the radii of the AdS and sphere
factors. As a result we would not expect  to get any corrections to the supersymmetry variation at 
the near 
horizon limit. The only new feature is that for small black holes we get supersymmetry enhancement
since higher order corrections stretch the horizon leading to $AdS_2\times S^3$ near horizon geometry.

In the following section we will use the ${\cal N}=2$ supercharges to study supersymmetry enhancement of small
black holes in ${\cal N}=4$ five dimensional supergravity.

\section{${\cal N}=4$ 5D black hole}

In the previous section we have shown that the small black hole in ${\cal N}=2$ supergravity in five dimensions
exhibits supersymmetry enhancement in the near horizon limit when the higher order corrections are added.
It is in contrast with what we have in leading order where it is known that only the large black hole exhibits
supersymmetry enhancement while the small one is singular without any supersymmetry doubling \cite{{Cham},{Townsend},{Sabra}}. 
 
In this section we would like to extend our study to small black holes in 5D ${\cal N}=4$ supergravity.
These solutions are $\frac{1}{2}$ BPS preserving eight supercharges and are singular at the tree level. Therefore 
taking the near horizon limit we would not expect to see supersymmetry doubling and the near horizon
geometry still preserves eight supercharges. In this section by making use of the fact that the
higher order corrections will stretch the horizon in such a way as to make the near horizon geometry 
$AdS_2\times S^3$, the supersymmetry enhancement emerges again. 

Of course to make the issue precise one first needs to show that there is a small black hole solution
in ${\cal N}=4$ supergravity in five dimensions in the presence of higher order corrections. In other words although the near horizon
information is useful, it is not enough to prove whether or not there is a solution interpolating between
the near horizon geometry and asymptotically flat space times and a priori it is not obvious whether the solution exists. 
So far, such a solution 
has not been found. Nevertheless there is an indirect evidence for the existence of such a solution.

An indication of the existence of a small black hole solution in the ${\cal N}=4$ case in the presence
 of higher order corrections would be if the five 
dimensional small black hole solution of the ${\cal N}=2$ obtained in \cite{Castro:2007hc} could indeed be embedded 
in the  ${\cal N}=4$ theory. The procedure 
is similar to the case for small black string studied in \cite{strominger}. The reason that the  embedding is 
possible is the fact
that if we regard the 5D small black hole solution to be the result of the reduction to 10 dimensional supergravity, because of the 
particular form of the charges and fields, the reduced background will not break the $Sp(4)$ R-symmetry of the 
${\cal N}=4$ model.
Therefore the supersymmetry
variation expressions are exactly the same as those in the previous section for ${\cal N}=2$ case. As a result the
supersymmetry enhancement works as in the ${\cal N}=2$ case studied in the previous section. The only difference is that
the index of the spinors $\varepsilon^i,\;\;i=1,2$ of the ${\cal N}=2$ model will now run from 1 to 4, $i=1,2,3,4$ for ${\cal N}=4$
case. In other words for the former case the spinors are in the {\bf 2} of $Sp(2)$ while for latter
case it is the {\bf 4} of $Sp(4)$. 

Therefore in the near horizon geometry of the small black hole of the ${\cal N}=4$ model where the geometry 
is $AdS_2\times S^3$, we get sixteen supercharges corresponding to
\bea
\varepsilon^i=\sqrt{\frac{r}{l}} \Omega\;\varepsilon_0^i,\;\;\;\;\;\;
\lambda^i=\frac{l}{2}(t-\frac{\gamma^{\hat{r}}}{r})\;\varepsilon^i,\;\;\;\;\;\;{\rm with}\;\;\;
\Omega=\sqrt{\frac{r}{l}}\;e^{\frac{1}{2}\gamma^{{\hat \psi}\hat{\phi}}\theta}\;  
e^{-\frac{1}{2}\gamma^{\hat{\theta}\hat{\phi}}(\psi+\phi)},
\eea
for $i=1,2,3,4$.

\subsection{Near Horizon superalgebra}

In this subsection we would like to construct the near horizon superalgebra for the ${\cal N}=4$ five
dimensional small black holes. As we have argued taking into account the 
higher order corrections will remove 
the singularity of the small black hole leading to $AdS_2\times S^3$ near horizon geometry where we get
supersymmetry enhancement.
Since we have a factor of $AdS_2$ one would expect to get a factor of $SO(2,1)$ in the near horizon
superalgebra. In our notation the corresponding isometry is generated by
\be
L_1=\frac{2}{l}\partial_t\ ,\;\;\;\;\;\;\;\;L_0=t\partial_t-r\partial_r\ ,\;\;\;\;\;\;\;\;
L_{-1}=\frac{l}{2}(r^{-2}+t^2)\partial_t-lrt\partial_r\ ,
\ee
satisfying $[L_m,L_n]=(m-n)L_{m+n}$, for $m,n=\pm 1,0$.

In order to find the near horizon superalgebra, one first needs to see how the AdS isometry acts on the 
supercharges. Using the explicit representation of the generators one finds
\be
L_0\lambda^i=\frac{1}{2}\lambda^i,\;\;\;\;\;\;L_0\varepsilon^i=-\frac{1}{2}\varepsilon^i,\;\;\;\;\;\;
L_1\lambda^i=\varepsilon^i,\;\;\;\;\;\;\;L_{-1}\varepsilon^i=-\lambda^i.
\ee
Here the action of generators is defined by the Lie derivative 
\be
{\cal L}_K\varepsilon^i=
(K^\mu {\cal D}_\mu+\frac{1}{4}\partial_\mu K_\nu \gamma^{\mu\nu})\varepsilon^i.
\ee
Therefore one may consider a correspondence between $\lambda^i$ and $\varepsilon^i$  and
 the $G_{-\frac{1}{2}}$ and $G_{\frac{1}{2}}$ modes
of supercurrent $G$, respectively. So that
\be
[L_m,G_r]=(\frac{m}{2}-r)G_{m+r}.
\ee
 
The next step is to do the same for $S^3$ part. In other words we will be looking for the action of $SO(4)$ generators on the
spinors. To study the action of the generators we use the fact that locally $SO(4)\approx SU(2)\times SU(2)$. In our notation
the generators of the two $SU(2)$'s are 
\bea
J^3=-\frac{i}{2}(\partial_\phi+\partial_\psi),&&\;\;\;\;\;\;J^{\pm}=\frac{1}{2}e^{\pm i(\psi+\phi)}(-i\partial_\theta
\pm\cot\theta\;\partial_\phi \mp\tan\theta\; \partial_\psi),\cr && \\
K^3=-\frac{i}{2}(\partial_\phi-\partial_\psi),&&\;\;\;\;\;\;K^{\pm}=\frac{1}{2}e^{\mp i(\psi-\phi)}(-i\partial_\theta
\pm\cot\theta\;\partial_\phi \pm \tan\theta\; \partial_\psi).\nonumber
\eea
On the other hand since $\gamma^{\hat{t}}$ and $\gamma^{\hat{\theta}\hat{\phi}}$ commute, we can always choose $\varepsilon^i_0$ such that
\be
\gamma^{\hat{\theta}\hat{\phi}}\varepsilon^i_0=\pm i\varepsilon^i_0,\;\;\;\;\;\;\;\;\;\;\gamma^{\hat{t}\hat{r}\hat{\psi}}
\varepsilon^i_0=\mp i\varepsilon^i_0.
\ee
With this definition we have
\be
J^3\varepsilon^i=\mp\frac{1}{2}\varepsilon^i,\;\;\;\;\;\;J^3\lambda^i=\mp\frac{1}{2}\lambda^i,\;\;\;\;\;\;K^3\varepsilon^i=0,\;\;\;\;\;
K^3\lambda^i=0.
\ee
Therefore the Killing spinor $\varepsilon^i$ and $\lambda^i$ are in the {\bf 2} representation of the first $SU(2)$ group generated by $J$ and are
neutral under the second $SU(2)$ group generated by $K$.

%Following \cite{strominger} 
Let us start with a constant spinor $\varepsilon_0$ such that 
$\gamma^{\hat{\theta}\hat{\phi}}\varepsilon_0=-i\varepsilon_0$. Then we can define
\be
\xi_+=\sqrt{\frac{r}{l}}\;e^{\frac{\theta}{2}\gamma^{\hat{\psi}\hat{\phi}}}e^{\frac{i}{2}(\psi+\phi)}\varepsilon_0,\;\;\;\;\;\;\;\;\;\;\;\;\;\;
\xi_-=\sqrt{\frac{r}{l}}\;e^{\frac{\theta}{2}\gamma^{\hat{\psi}\hat{\phi}}}e^{-\frac{i}{2}(\psi+\phi)}\gamma^{\hat{\psi}\hat{\theta}}\varepsilon_0,
\ee
and normalize to $\varepsilon_0^\dagger \varepsilon_0=1$. It is easy to verify that
\be
J^3\xi_\pm=\pm\xi_\pm,\;\;\;\;\;\;\;J^{\pm}\xi_\pm=0,\;\;\;\;\;\;\;J^{\pm}\xi_\mp=\xi_{\pm},
\ee
and therefore $\xi$ is in the {\bf 2} of the first $SU(2)$ group. Using this notation one may express the Killing spinors, 
$\varepsilon^I$, corresponding to the supercharges as follows \cite{strominger}
\bea\label{aaa}
\varepsilon^1=\left(\begin{array}{c}\xi_+ \\
i\xi_-\\ 0\\ 0\end{array}\right),\;\;\;\;\varepsilon^2=\left(\begin{array}{c}-i\xi_+ \\
-\xi_-\\ 0\\ 0 \end{array}\right),\;\;\;\;\varepsilon^3=\left(\begin{array}{c}\xi_- \\
-i\xi_+\\ 0 \\ 0 \end{array}\right),\;\;\;\;\varepsilon^4=\left(\begin{array}{c}i\xi_- \\
-\xi_+\\ 0 \\ 0 \end{array}\right),&&
\cr &&\\
\varepsilon^5=\left(\begin{array}{c}0\\0\\ \xi_+ \\
i\xi_-\end{array}\right),\;\;\;\;\varepsilon^6=\left(\begin{array}{c}0\\0\\ -i\xi_+ \\
-\xi_-\end{array}\right),\;\;\;\;\varepsilon^7=\left(\begin{array}{c}0\\0\\ \xi_- \\
-i\xi_+\end{array}\right),\;\;\;\;\varepsilon^8=\left(\begin{array}{c}0\\0\\ i\xi_- \\
-\xi_+\end{array}\right),&&\nonumber
\eea
which correspond to $G_{\frac{1}{2}}^I,\;I=1,\cdots,8$. Similarly $G_{-\frac{1}{2}}^I$ corresponds to
$\lambda^I$ given by
\be
\lambda^I=\frac{l}{2}(t-\frac{\gamma^{\hat{r}}}{r})\varepsilon^I,\;\;\;\;\;\;\;\;\;\;
{\rm for}\;\;I=1,\cdots, 8.
\label{super2}
\ee
Here each $\lambda^I$ or $\varepsilon^I$ transforms as the {\bf 4} of $Sp(4)$.  

To complete the near horizon superalgebra we need to compute the anticommutators of supercharges. To do this we use
the supersymmetry transformations of the five dimensional supergravity given by \cite{Hanaki:2006pj}
\bea
\{G^I_r,G^J_s\}&=&l\Omega_{ij}\left[(\bar{\varepsilon}^I_r)^i\gamma^\mu ({\varepsilon}^J_s)^j+
(\bar{\varepsilon}^J_s)^i\gamma^\mu({\varepsilon}^I_r)^j\right]\partial_\mu\cr &&\cr
&+&\left[(\bar{\varepsilon}^I_r)_i\gamma^{\hat{r}\hat{t}} ({\varepsilon}^J_s)^j+
(\bar{\varepsilon}^J_s)_i\gamma^{\hat{r}\hat{t}} ({\varepsilon}^I_r)^j\right],
\label{com}
\eea
where $\Omega_{ij}$ is a symplectic matrix which raises and lowers indices as $\varepsilon_i=
\Omega_{ij}\varepsilon^j$. We choose a basis in which $\Omega_{12}=\Omega_{34}=1$. 
Plugging the supercharges (\ref{aaa}) and (\ref{super2}) into (\ref{com}) we get the anticommutators of the supercharges, e.g.,
\be
\{G_{\pm\frac{1}{2}}^I,G_{\pm\frac{1}{2}}^J\}=2\delta^{IJ}L_{\pm 1},
\ee 
and 
\be
\{G_{\frac{1}{2}}^I,G_{-\frac{1}{2}}^J\}=\left(\begin{array}{cccc} 2L_0 & 2iJ^3+iA_3 & 2iJ^2+i A_1 &-2iJ^1+iA_2\\
-2iJ^3-iA_3&2L_0 & -2iJ^1-i A_2 &-2iJ^2+iA_1\\
-2iJ^2-iA_1 & 2iJ^2+i A_1 &2L_0 &-2iJ^3-iA_3\\
2iJ^1-iA_2 & 2iJ^2-i A_1 &2iJ^3+iA_3 & 2L_0\end{array}\right).
\ee 
for $I,J=1,2,3,4$. We have computed the other anticommutator relations and we can summarize the entire superalgebra as follows
\bea
&&\{G^I_r,G^J_s\}=2\delta^{IJ}L_{r+s}+(r-s)(M_{a})^{IJ}J^a+(r-s)(N_A)^{IJ}T^A,\cr &&\cr
&&[L_m,G_r^I]=(\frac{m}{2}-r)G_{m+r}^I,\;\;\;\;\;\;[L_m,L_n]=(m-n)L_{m+n}\cr &&\cr
&&[T^A,G^I_r]=(N^{A})^{IJ} G_r^J,\;\;\;\;\;\;\;\;\;\;\;\;\;[J^a,G_r^I]=(M^{a})^{IJ}G^J_r,
\eea
where $T^A$ are the generators of $Sp(4)$ parameterized by $T^A=\{A_\alpha,B_\alpha,C_\alpha,C_0\}$ 
\be
A_\alpha=\left(\begin{array}{cc}\sigma^\alpha & 0\\ 0& 0\end{array}\right), \;\;\;\;B_\alpha=\left(\begin{array}{cc}0 & 0\\ 
0& \sigma^\alpha\end{array}\right),\;\;\;\;C_\alpha=\left(\begin{array}{cc}0& \delta^\alpha  \\ 
{\delta^\alpha}^\dagger& 0\end{array}\right),\;\;\;\;C_0=\left(\begin{array}{cc}0& \frac{i}{2}  \\ 
\frac{-i}{2}& 0\end{array}\right).
\ee
Here $\delta^\alpha=\frac{1}{2}(\sigma^1,i\sigma^2,\sigma^3)$ with $\sigma^\alpha$ being  the Pauli matrices.
$M_a$ and $N_A$ are the representation matrices for $SU(2)$ and $Sp(4)$, respectively.

This is, indeed, the commutation relations of the supergroup $OSp(4^*|4)$ which also
appeared for the near horizon of ${\cal N}=4$ five dimensional small black string \cite{strominger}.
We note, however, that in our case we have another $SU(2)$ coming from the $SO(4)$ isometry of the $S^3$ factor. 
Of course the generators of this $SU(2)$
commute with all the other generators. So the algebra is 
\be
OSp(4^*|4)\times SU(2).
\ee 
We note that the bosonic part of the global supergroup $OSp(4^*|4)\times SU(2)$ is $SL(2)\times SU(2)\times SU(2)\times Sp(4)$
while the isometry of the near horizon geometry is $SL(2)\times SO(4)$. Therefore there is an extra $Sp(4)$ symmetry which can not
be  geometrically realized. Following \cite{strominger} we may identify this symmetry with R-symmetry of ${\cal N}=4$ supergravity
in five dimensions.

\section{${\cal N}=2$ revisited; From ${\cal N}=4$ to ${\cal N}=2$}

As we have seen the supersymmetry enhancement depends crucially on the geometry of the
near horizon limit. The appearance of $AdS_2$ factor is essential in getting supersymmetry
doubling. Nevertheless we note that the
whole background will be fixed not only by the metric but also by other fields, such as
gauge field, scalar fields, and so on. So one might expect that the actual number of supercharges 
present in the near horizon geometry  depends on the other fields too.

This being the case in general, in the ${\cal N}=2$ case the way supersymmetry 
enhancement appeared in the previous sections, with the higher order corrections, seems to be special and 
blind to the other
fields. Therefore a priori it is not obvious how the other fields affect the supersymmetry. 
In particular it is not clear how the supersymmetry distinguishes between the large and the small
black holes or black strings in ${\cal N}=2$ theory considering  that in both cases
eight supercharges are preserved in the presence of higher order corrections.

Note however, that this question does not arise for ${\cal N}=4$ case, as the large
black holes/strings are $\frac{1}{4}$ BPS while the small ones are $\frac{1}{2}$ BPS. Thus
taking the near horizon limit they lead to AdS geometries with different number of supercharges; 
eight and sixteen 
supercharges, respectively. 
The situation with the supergroups of the  ${\cal N}=4$ theory are summarized in 
the table (\ref{Global1}).
\begin{table}
  \centering
  \begin{tabular}{|c|c|c|c|}
  \hline
\textbf{Object}&  \textbf{Bosonic Symmetry} & \textbf{supercharges} & \textbf{ Supergroup}    \\
\hline
\text{L-black string} & $SL(2)\times SU(2)$   & 8 & $SU(1,1|2)$ \\
\hline
\text{L-black bole}   & $SL(2)\times SO(4)$  &  8 & $SU(1,1|2)\times SU(2)$\\
\hline
\text{S-black string}   & $SL(2)\times SU(2)\times Sp(4)$  & 16 & $OSp(4^*|4)$ \\
\hline
\text{S-black hole} & $SL(2)\times SO(4)\times Sp(4)$ &  16 & $OSp(4^{*}|4)\times SU(2)$\\
\hline
\end{tabular}
\caption{Global supergroup of near horizon geometry of small (S) and large (L) black holes and strings in the ${\cal N}=4$ 
five dimensional supergravity.}
\label{Global1}
\end{table}
For ${\cal N}=2$ model the situation is quite different. As we already mentioned the problem appears because both the small and large
black holes (strings) are $\frac{1}{2}$ BPS and in the near horizon, when higher order corrections are taken into account,
 preserve the same number of eight supercharges. Of course in leading order there is a big difference between these two cases and in fact only
the large black holes (strings) would exhibit supersymmetry doubling while for small ones the geometry is singular and at near horizon
we will still have four supercharges. 
We would like to pose the question of how to distinguish between the small and large black holes for the higher
order corrected action. 
To answer to this question we will resort to the ${\cal N}=4$ model studied in the last section (or
the one considered in \cite{strominger})  by a process of reduction of the number of supersymmetries.

To get the ${\cal N}=2$ model from the ${\cal N}=4$ theory one may follow two different routes
\footnote{We would like to thank Joshua M. Lapan for discussions on this point.}: Starting from
a small black hole (string) solution in ${\cal N}=4$ theory,  
we can either add some matter fields or we can simply truncate some supercharges, ending up 
with a solution in ${\cal N}=2$ theory. Depending on which route we choose we get either a large or a small 
black hole (string).
Of course since we do not have an explicit solution for small black holes (strings) in the presence of $R^2$
corrections (see however \cite{Castro:2007hc}) in general it is difficult to do this reduction explicitly.
Nevertheless one may proceed for the near horizon geometry.  To be specific consider 
the small black string in ${\cal N}=4$ theory and try to find the near horizon superalgebra when
the reduction to the ${\cal N}=2$ case is carried out.

The small black string in five dimensions from the Heterotic string theory point of view, corresponds
to a fundamental string living on $R^{1,4}\times T^5$. Adding matter fields from string theory point of view
corresponds to turning on some other charges. In particular we can add a set of NS5-branes
wrapped on the $T^4$, with its fifth direction along the fundamental string. In this case
the background in the near horizon geometry preserves just eight supercharges and indeed this will
turn out to be a large black string with near horizon supergroup $SU(1,1|2)$ as in table (\ref{Global1}).
Now truncating the solution to ${\cal N}=2$ supergravity we will end up with near horizon
geometry of large black strings in the ${\cal N}=2$ theory. The near horizon global symmetry will still 
remain the same, i.e.
 $SU(1,1|2)$. Note that the near horizon supergroup can be directly obtained from ${\cal N}=2$ theory too
using the method we used in the previous section.
(see for example \cite{Townsend}). 

On the other hand, one could start from a small black string in ${\cal N}=4$ theory and just throw away 
half of the supercharges. Actually this is the reverse procedure we used to get our ${\cal N}=4$ solution
from ${\cal N}=2$ solution. Doing so we end up with a small black string in ${\cal N}=2$ theory. It will preserve 
eight supercharges corresponding to
\bea
\varepsilon^1=\left(\begin{array}{c}\xi_+ \\
i\xi_-\end{array}\right),\;\;\;\;\varepsilon^2=\left(\begin{array}{c}-i\xi_+ \\
-\xi_-\end{array}\right),\;\;\;\;\varepsilon^3=\left(\begin{array}{c}\xi_- \\
-i\xi_+ \end{array}\right),\;\;\;\;\varepsilon^4=\left(\begin{array}{c}i\xi_- \\
-\xi_+\end{array}\right),
\label{yyyy}
\eea
and 
\be
\lambda^I=\frac{l}{2}(t-\frac{\gamma^{\hat{r}}}{r})\varepsilon^I,\;\;\;\;\;\;\;\;\;\;
{\rm for}\;\;I=1,2,3,4.
\ee
In this case the $Sp(4)$ R-symmetry will break to $Sp(2)$ and it is easy to show that the
bosonic part of the symmetry will be $SL(2)\times SU(2)\times Sp(2)$. Searching in the literature (for example see \cite{Ant}) 
we find that there is, indeed, a supergroup with this bosonic part and supporting eight supercharges which is
$D(2,1;\alpha)$\footnote{We would like to thank A. Giveon for a comment on this point.}. 
The parameter $0<\alpha\leq 1$  is a relative weight of $SU(2)$ and $Sp(2)$. Clearly this parameter is
not determined by a knowledge of the bosonic symmetry. Nevertheless using the direct computations as carried out  in the previous 
section we will be able to find $\alpha$.
The procedure is the same for small black holes. In particular, using our notations in the previous section and also the spinors
(\ref{yyyy}), the global algebra 
is as follows
\bea
&&\{G^I_r,G^J_s\}=2\delta^{IJ}L_{r+s}+(r-s)(M_{a})^{IJ}J^a+(r-s)(N_A)^{IJ}T^A,\cr &&\cr
&&[L_m,G_r^I]=(\frac{m}{2}-r)G_{m+r}^I,\;\;\;\;\;\;[L_m,L_n]=(m-n)L_{m+n}\cr &&\cr
&&[T^A,G^I_r]=(N^{A})^{IJ} G_r^J,\;\;\;\;\;\;\;\;\;\;\;\;\;[J^a,G_r^I]=(M^{a})^{IJ}G^J_r.
\eea
Here $T^A$ are the generators of $Sp(2)$ given by the Pauli matrices and $N_A$ is representation matrix
for $Sp(2)$. This is, indeed, the commutation relations of $Osp(4^*|2)=D(2,1;1)$, i.e. $\alpha=1$.
With an extra $SU(2)$ coming from $SO(4)$ generated by $J$ we get $Osp(4^*|2)\times SU(2)$ as
the global near horizon supergroup of the small black hole in ${\cal N}=2$ supergravity in five
dimensions. While for large black hole in this model the global near horizon superalgebra is $SU(1,1|2)\times SU(2)$.
Therefore as we see the near horizon supergroup of small and large 
black strings/holes in ${\cal N}=2$ does distinguish between being small or large.  The final results are    
summarized in table (\ref{Global2}).
\begin{table}
  \centering
  \begin{tabular}{|c|c|c|c|}
  \hline
\textbf{Object}&  \textbf{Bosonic Symmetry} & \textbf{supercharges} & \textbf{ Supergroup}    \\
\hline
\text{L-black string} & $SL(2)\times SU(2)$   & 8 & $SU(1,1|2)$ \\
\hline
\text{L-black bole}   & $SL(2)\times SO(4)$  &  8 & $SU(1,1|2)\times SU(2)$\\
\hline
\text{S-black string}   & $SL(2)\times SU(2)\times Sp(2)$  & 8 & $OSp(4^*|2)$ \\
\hline
\text{S-black hole} & $SL(2)\times SO(4)\times Sp(2)$ &  8 & $OSp(4^{*}|2)\times SU(2)$\\
\hline
\end{tabular}
\caption{Global supergroup of near horizon geometry of small (S) and large (L) black holes and black string in ${\cal N}=2$ five dimensional supergravity.}
\label{Global2}
\end{table}
We observe that as in the ${\cal N}=4$ case the small black hole has an extra factor of $Sp(2)$ symmetry which
cannot be geometrically realized from our near horizon geometry. Nevertheless, following \cite{strominger} we would like
to identify this extra symmetry with the R-symmetry of ${\cal N}=2$ supergravity in five dimensions.

\section{Conclusions}

In this paper we have shown that the supersymmetry enhancement can also occur for small black holes 
in ${\cal N}=2$ and ${\cal N}=4$ five dimensional supergravity. For large black holes this effect has known 
for about a decade \cite{Cham}. What makes the large black holes easier to handle is that 
at leading order the large black hole has near horizon geometry of the form $AdS_2\times S^3$ while
the small black hole is singular. But with inclusion  of higher order corrections in the action 
both large and small black hole solutions become smooth with $AdS_2\times S^3$ near horizon geometry. 
It is then possible in both cases
to show supersymmetry enhancement to eight supercharges.

However We have argued  that the corresponding global near horizon supergroups are
different; for the large black hole it is $SU(1,1|2)\times SU(2)$ \cite{Townsend} while for the 
small black hole it turns out to be $OSp(4^*|2)\times SU(2)$.

An immediate puzzle we face is that in the small black hole, unlike the large one, there is no  
one to one correspondence between the isometry of the near horizon geometry and the bosonic part of the
supergroup. In particular there is an extra $Sp(2)$ factor in the corresponding supergroup. Nevertheless following
\cite{strominger} we note that there is a novel way to interpret this extra symmetry: It can be
interpreted as the R-symmetry of ${\cal N}=2$ five dimensional supergravity.  
If this interpretation is  correct it is not clear to us why for the large black holes this factor is
absent.

Another comment we would like to make is that whenever we have $AdS_2$ or $AdS_3$ factor the superisometry 
must have an affine extension containing a Virasoro algebra \cite{{brown},{Strominger:1998yg}}. 
Therefore the supergroup we have obtained for small black hole, $OSp(4^*|2)\times SU(2)$, is expected to be
the zero mode algebra of a corresponding unknown affine algebra.

The above conclusions can also be made for small and large black strings in ${\cal N}=2$ five 
dimensional supergravity.  The corresponding near horizon supergroup is $OSp(4^*|2)$ for small
and $SU(1,1|2)$ for large black strings. Since the near horizon supergroup for small and large
black strings are different, it would be interesting to understand how this will affect the
properties of the corresponding 2D conformal field theory holographically dual to these backgrounds.  
Small black string of ${\cal N}=4$ supergravity in five dimensions and its holographic dual  have recently 
been studied in \cite{{strominger},{Dabh},{johnson},{kraus}} (see also \cite{{giveon},{HU}}).

For the ${\cal N}=4$ theory we have shown that the near horizon geometry of small black holes preserves sixteen
supercharges. Moreover by making use of the Killing spinor analysis we have shown that the
global near horizon supergroup is $OSp(4^*|4)\times SU(2)$. Again in this case we have 
an extra $Sp(4)$ factor which cannot be realized geometrically as the part of the isometry of the
near horizon geometry. But, it may be interpreted as the R-symmetry of ${\cal N}=4$ five
dimensional supergravity. 
There is no confusion however  between large and small black holes in this case as they preserve different 
number of supercharges. 

The next step would be to look for an affine extension of the supergroup. As mentioned in \cite{strominger} there are 
no linear superconformal algebras with more than eight supercharges. Nevertheless if we relax the 
linearity condition for the algebra there is non-linear affine algebra $\widehat{OSp(4^*|4)}$ which contains
the $OSp(4^*|4)$ in the large central charge limit . We note, however, that even though
this affine algebra contains the part we are interested in ( in a specific limit), it is not clear if it is 
physically acceptable, e.g., as also noted in \cite{kraus} this algebra does not have any unitary representations. 
Moreover it is not clear how to incorporate
the extra $Sp(4)$ factor in the affine structure.

\vspace*{1cm}

{\bf Acknowledgments}

We would like to thank H. Arfaei, J .P. Gauntlett, J. M. Lapan and F. Larsen for useful discussions. M. A. and F. A. would also like to thank CERN theory division  
and S.M. thanks IPM, Tehran for hospitality where 
parts of this work were done. This work is supported in part by 
Iranian TWAS chapter at ISMO.

\end{document}